\documentclass{elsart}

\usepackage{graphicx}

\begin{document}

\begin{frontmatter}

\title{Comment on ``Estimating $1/f^\alpha$ scaling exponent from short time series"}

\author{P.\ F.\ G\'ora}
\ead{gora@if.uj.edu.pl}
\address{M.\ Smoluchowski Institute of Physics, Jagellonian University, Reymonta~4, 
30--059~Krak\'ow, Poland}

\begin{abstract}
A recently proposed method (O. Miramontes, P. Rohani, Physica D 166 (2002) 147) 
for estimating the scaling exponent in very short time series may give wrong results, 
especially in case of undersampled data.
\end{abstract}

\begin{keyword}
Time--series \sep Noise parameter estimation
\PACS 05.45.Tp \sep 87.10.+e
\end{keyword}

\end{frontmatter}

Estimating the scaling exponent in the power spectrum of a signal is one of the
most important tasks in the analysis of real--world temporal sequences. 
The point 
of time series analysis is not making a statement about the signal in question, 
but getting insight of the (unknown) mechanism responsible for generating the 
signal. One may say that the time series is usually not very interesting,
but its generating mechanism is.
The scaling exponent is one of the quantities that are calculated in order to
get a better understanding of this mechanism. Many techniques
have been developed for this job; see Ref.~\cite{kaplan} for a recent review.
It is particularly difficult to estimate the scaling exponent if the time
series is very short, which is frequent for biological, medical
and ecological data. Indeed, it is not clear whether the notion of the scaling exponent
has any well--defined meaning in this case.

Short time series are usually characterized by a 
large sampling time, $T$. This is easy to understand: The series is short because
it takes long time to gather the data, and it takes long time to gather the data because
the sampling time is large and cannot be easily shortened.
In a time series only the frequencies up to the Nyquist frequency $f_{\mathrm{N}}=
1/(2T)$ are present \cite{NR}. Any 
possible power scaling that is discovered within the series is necessarily restricted 
to the Nyquist interval $[0,f_{\mathrm{N}}]$. This is all the data can tell --- 
there are no scientific 
grounds to claim that the scaling extends beyond the Nyquist frequency. 
Because $T$ is large, $f_{\mathrm{N}}$ is small. Therefore,
attributing the power scaling, if one is detected in such a series, to noise is misleading 
at best, as in a true
noise the power scaling extends to many (theoretically to infinitely many) orders
of magnitude.

When dealing with a short time series, one is also faced with another, even more acute
problem: the Nyquist interval  is sparsely covered. There is
little one can tell about the power density at frequencies that are not multiplies
of $1/(2NT)$, where $N$ is the lenght of the series,
and it is not clear whether the very notion of a power scaling can be used
in such situations. Any attempts to interpolate between frequencies directly
accessible form the data merely scale down the behaviour observed for these 
frequencies to those that are not directly accessible. This approach
may work only if one knows \textit{a priori\/} that the process used to generate the data 
is governed by a power law.

A method published recently in Ref.~\cite{paper} suffers from all these deficiencies.
The method, named ``multiple segmenting method'' (MSM), consists of two steps:
First, one calculates scaling exponents for each segment of length $2^s$, $s\ge3$, 
and calculates the average scaling exponents for each~$s$; denote these averages by 
$g(n)$ with $n=2^s$. 
In this step pseudo--replicates of the original data are created. These 
pseudo-replicates have the same short--time, or large--frequency, correlations as
the original series, and the process of replicating increases their statistical 
significance. However, any accidental correlations among the data are enhanced
as well. Second, the relation

\begin{equation}\label{sqrtn}
g(n)=a+\frac{b}{\sqrt{n}}
\end{equation}

\noindent is fitted to the averages calculated in the first step, and the 
scaling exponent for the whole series is calculated from (\ref{sqrtn}) by
using results of the fit and $n=N$, where $N$ is the actual length of the
time series. The equation (\ref{sqrtn}) has not 
been derived in any rigorous way, only guessed from the example data.

\begin{figure}
\begin{center}
\includegraphics[scale=0.63]{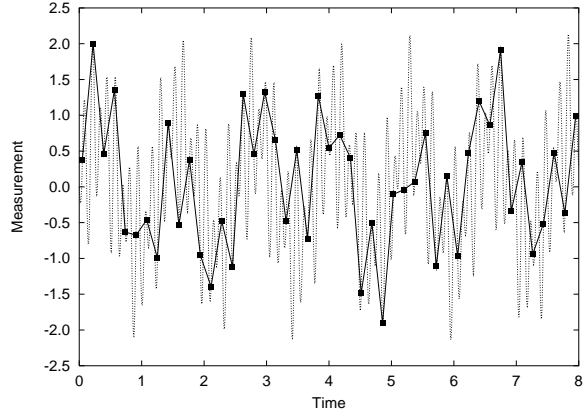}
\end{center}
\caption{A quasiperiodic function (\protect\ref{quasi}) (thin line) and
a time series resulting from undersampling the function (boxes).}\label{series-quasi}
\end{figure}

\begin{figure}
\begin{center}
\includegraphics[scale=0.63]{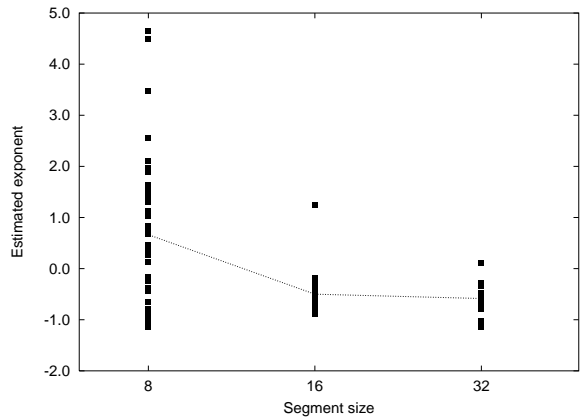}
\end{center}
\caption{MSM applied to the data from Fig.~\protect\ref{series-quasi}. The
thin line connects the averages for $n=8$, 16, 32.}\label{msm1}
\end{figure}

The authors of Ref.~\cite{paper} claim that MSM gives correct results for time series as short
as $N=47$ terms, but offer little proof to this claim. They present a number
of examples with artificially created time series of moderate length $N=400$,
almost an order of magnitude larger than 47. The time series used had,
by construction, built--in power scaling, and so did all their subseries.
In these cases the MSM method discovered the scaling exponents
that we beforehand knew were present in the data. Two out of four real--world 
examples presented had time series even longer ($N=2000$ and $N=1024$, respectively), 
where the power--type behaviour could be detected and the corresponding scaling
exponents determined by conventional methods. There were only two really
short ($N=47$) time series of unknown properties discussed. The MSM method,
when applied to these two series, gave some numbers as the scaling exponents,
but it is not clear whether these numbers are at all meaningful. 

\begin{table}
\begin{center}
\begin{tabular}{r||r|r|r}
$n$&$g_1(n)$&$g_2(n)$&$g_3(n)$\\
\hline
 8&$ 0.66$&$-0.26$&$-1.25$\\
16&$-0.50$&$-0.33$&$-1.52$\\
32&$-0.59$&$-0.16$&$-1.05$\\
47&$-0.98$&\multicolumn1{|c|}{---}&\multicolumn1{|c}{---}
\end{tabular}
\end{center}
\caption{Results of MSM method applied to the three example time
series: $g_1(n)$ --- the undersampled quasiperiodic function
from Fig.~\protect\ref{series-quasi}, $g_2(n)$ --- the undersampled AR(1)
process from Fig.~\protect\ref{ARfig}, $g_3(n)$ --- first 47 terms of the
full AR(1) series.
The last row, second column, shows the result of applying the formula 
(\protect\ref{sqrtn}).}
\label{tabela1}
\end{table}

In what follows we present examples of short time series where the MSM
method, when blindfoldedly applied, gives clearly misleading results.

\begin{figure}
\begin{center}
\includegraphics[scale=0.63]{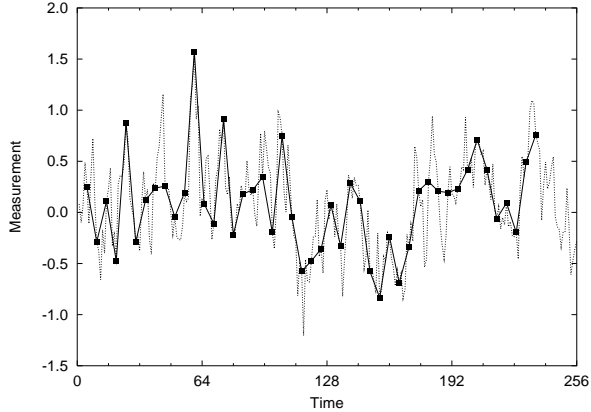}
\end{center}
\caption{A realization of the AR(1) process (\protect\ref{AReq}) (thin line)
and an undersampled substring of this signal (boxes). The undersampled series
is our second example time series, and the first 47 terms of the series used
to draw the thin line is the third one.}\label{ARfig}
\end{figure}

As a first example, we take a quasiperiodic function

\begin{equation}\label{quasi}
x(t) = \sum\limits_{i=1}^3 a_i \sin(\omega_i t),
\end{equation}

\noindent with $a_1=0.8$, $\omega_1=5$, $a_2=0.6$, $\omega_2=20\sqrt{3}$,
$a_3=-0.8$, $\omega_3=30\sqrt{3}$, sampled with a time step $T=7.9/46\simeq
0.17174$ to give $N=47$. This time series is clearly undersampled, and many 
of the high--frequency features of the function (\ref{quasi}) are lost, 
cf.~Fig~\ref{series-quasi}. When we apply the MSM method to this time series, 
we obtain a scaling exponent of about $-0.98$, cf.~Fig.~\ref{msm1} and 
Table~\ref{tabela1}. This result, if taken at its face value, 
indicates that the time series is governed by a flicker (or pink) noise, 
which is obviously untrue: the microscopic mechanism responsible for this 
time series is a regular, quasiperiodic function, albeit undersampled. 

\begin{figure}
\begin{center}
\includegraphics[scale=0.63]{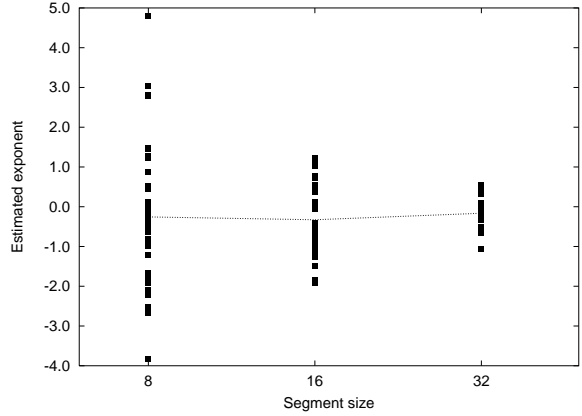}
\end{center}
\caption{Results of the MSM method applied to the second example time series.
The thin line connects the averages for $n=8$, 16, 32.}
\label{msmAR5}
\end{figure}

\begin{figure}
\begin{center}
\includegraphics[scale=0.63]{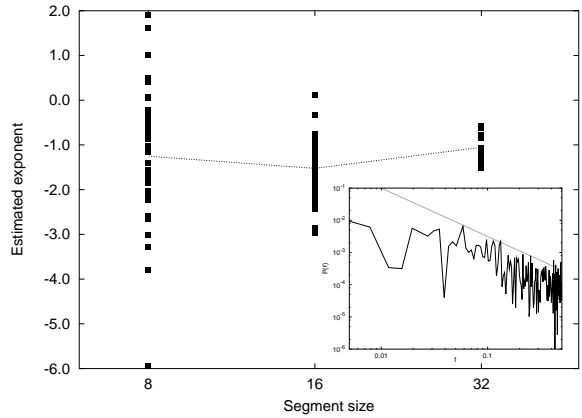}
\end{center}
\caption{Results of the MSM method applied to the third example time series
(47 terms). The thin line connects the averages for $n=8$, 16, 32. The inset 
shows the power spectrum calculated from the whole 256 terms used to generate 
the data. The line with the slope $-1.5$ is not a rigorous fit --- it is meant 
as a guide to the eye only.}\label{msmAR1}
\end{figure}

We used a realization of the Markovian AR(1) process \cite{box}

\begin{equation}\label{AReq}
X_{n+1} = \beta X_n + (1-\beta)\eta_n\,,
\end{equation}

\noindent with $\beta=0.69$ like in \cite{paper}, to generate two next example
time series. $\eta_n$ is a Gaussian white noise, generated numerically by means 
of an algorithm published in \cite{fernandez}. We generated 256 terms of the 
series (\ref{AReq}), and then constructed our second example time series
by taking every fifth term of the original series up to $N=47$ 
($X_5$, $X_{10}$,\dots,$X_{235}$). Finally, the
first consecutive 47 terms of the original series ($X_1$, $X_2$,\dots,$X_{47}$) 
were used as the third
example. The output of the MSM method is presented in Figs.~\ref{msmAR5} 
and~\ref{msmAR1} and in Table~\ref{tabela1}. Note that the function 
(\ref{sqrtn}) does not seem
to be a reasonable fit to these data, and if this fit fails, MSM does not
offer any other means to systematically determine the scaling exponent
for the whole series. For the second example the method indicates the scaling
exponent in the range $[-0.33,-0.16]$, i.e. it recognizes the time series
as being close to the white noise, which is again not true. On the other hand,
for the third example the method predicts the scaling exponent in the range 
$[-1.52,-1.05]$, which is close (especially the lowest value) to a rough estimate
for the scaling exponent for the whole 256 term time series. Note that this
time series was, by construction, characterized by a power--type behaviour.
The relative success of MSM in this particular case strengthens our point 
that \textit{if\/} a power law is present in the data, scaling it down to
frequencies not observed in the data may not hurt much, but a short time
series does not tell \textit{whether\/} such a power law is present.

One may argue that the MSM method failed for the undersampled time series only.
However, undersampling can be quite common in practice.
When dealing with real--world time series generated by a mechanism
whose details remain unknown, we never know whether we choose a correct
sampling time. For instance, in the bacterial population data analysed in
Ref.~\cite{paper} the sampling time was as long as a week, while the true
dynamics of the population might have been governed by processes with a much
lower characteristic time, like daily changes in sunlight or temperature.
Similarly, in many medical studies levels of drugs in subjects are determined
on a daily basis, while physiological processes responsible for the drug absorption
and decay usually proceed much faster. 
There are many other situations in which data are collected not in a controlled
laboratory environment, but come out of real--world processes, where adjusting
the sampling time may be impossible for various ``technical" reasons.
In case of undersamplig, the power spectrum beyond the Nyquist frequency
is aliased into the  Nyquist interval, distorting the available
power spectrum \cite{NR}. Examples presented above show that 
the MSM method, when applied to such data, may give wildly wrong results.

One needs to know some important details of the underlying process in order
to estimate its parameters from a very short time series.
Otherwise idiosyncrasies of the data are mistakenly taken for true 
correlations and are enhanced in the process of 
pseudo--replicating the time series. This may lead to a failure of the MSM
method. This method appears to work tolerably well for long time series,
but in this case it is inferior to conventional techniques
due to its large computational costs. In our opinion, the usefulness of
the MSM method is thus limited to the rare case when one knows for sure
that the process used to generate the time series is governed by a power law
with an unknown exponent,
but the time series available is very short and cannot be easily extended
and only a rough estimate of the scaling exponent is required.
In other situations the MSM method is not recommended.

I would like to thank Dr.\ Adam Kleczkowski from Cambridge University
for his helpful comments.

\end{document}